\documentclass[aps,showpacs,tightenlines,twocolumn,nofootinbib,nobibnotes,superscriptaddress]{revtex4-1}
\usepackage{amsmath,amssymb,amsfonts,bm}
\usepackage{graphicx}
\usepackage{epstopdf}
\usepackage{dcolumn}
\usepackage{mathrsfs}
\usepackage{tgtermes}
\usepackage[colorlinks=true,linkcolor=red,citecolor=blue, urlcolor=blue,bookmarks=false]{hyperref}
\bibpunct{[}{]}{,}{n}{}{}
\usepackage{verbatim}
\usepackage{amsthm,amsmath,amssymb}
\usepackage{mathrsfs}
\usepackage{booktabs}
\usepackage{multirow}

\begin{document}
\title{Five-dimensional Floquet topological semimetals with emergent Yang monopoles and linked Weyl surfaces}
\date{\today}
\author{Zheng-Rong Liu}
\affiliation{Department of Physics, Hubei University, Wuhan 430062, China}
\author{Rui Chen}\email{chenr@hubu.edu.cn}
\affiliation{Department of Physics, Hubei University, Wuhan 430062, China}
\author{Bin Zhou}\email{binzhou@hubu.edu.cn}
\affiliation{Department of Physics, Hubei University, Wuhan 430062, China}
\affiliation{Key Laboratory of Intelligent Sensing System and Security of Ministry of Education, Hubei University, Wuhan 430062, China}

\begin{abstract}
Recently, Floquet topological matter has attracted significant attention for its potential to reveal novel topological phases inaccessible in static systems. In this paper, we investigate the effect of a time-periodic driving on the five-dimensional (5D) normal insulators. We show that the time-periodic driving can induce a topological phase transition from a \textbf{TP} (time reversal combined with space inversion) symmetry-preserving normal insulator to a 5D Floquet topological semimetal with emergent Yang monopoles characterized by the second Chern number. Additionally, we show that this time-periodic driving can also lead to a topological phase transition from a \textbf{TP} symmetry-breaking normal insulator to a 5D Floquet topological semimetal with a Hopf link formed by Weyl surfaces. Besides, further increasing the strength of the time-periodic driving, the two 5D Floquet topological semimetal phases are transformed into the 5D Floquet Chern insulators. Our paper is helpful for future studies in higher dimensional Floquet systems.
\end{abstract}

\maketitle

\section{Introduction}
Topological semimetals are a class of gapless topological matter that have drawn significant research interest both theoretically and experimentally~\cite{Burkov_2016, Yan_2017, RevModPhys.90.015001, RevModPhys.93.025002}. Topological semimetals not only exhibit a rich variety of bulk and surface transport phenomena but also act as intermediate states during topological phase transitions between distinct gapped topological materials~\cite{NIELSEN1983389, 10.1098/rspa.1984.0023, PhysRevLett.93.206602, 10.1088/1367-2630/9/9/356, PhysRevB.83.205101, doi:10.1126/sciadv.1602680, PhysRevB.108.195306, PhysRevB.108.085135}. In three dimensions, topological semimetals are classified as Dirac semimetals~\cite{doi:10.1126/science.1256742, PhysRevB.88.125427, Neupane_2014, Jeon_2014, PhysRevLett.107.186806, PhysRevLett.115.036807, PhysRevLett.131.096901, PhysRevLett.131.186302, PhysRevB.85.195320, PhysRevLett.108.140405, PhysRevLett.113.027603, PhysRevLett.127.066801, PhysRevB.108.L041104}, Weyl semimetals~\cite{Weyl_1929, PhysRevB.86.115133, PhysRevB.87.235306, PhysRevB.88.104412, PhysRevB.90.155316, doi:10.1126/science.aaa9297, PhysRevX.5.011029, Huang_2015, PhysRevX.5.031013, Balents_2011, PhysRevLett.123.065501, PhysRevB.101.235119, Yang_2023, PhysRevResearch.5.L022013, PhysRevB.108.235211, PhysRevLett.133.096601, Lu_2024}, and nodal-line semimetals~\cite{PhysRevB.84.235126, PhysRevB.92.045108, PhysRevLett.115.036807, ZHOU2022, PhysRevB.109.195421, PhysRevB.105.155102}. In Weyl semimetals, non-degenerate valence and conduction bands touch each other at zero-dimensional (0D) Weyl points. Weyl points appear in pairs and exhibit opposite chiralities~\cite{PhysRevB.83.205101, NIELSEN1983389, Balents_2011}, with the chirality determined by the first Chern number $C_{1}=\frac{1}{2\pi}\oint_{S^{2}}F d^{2}\textbf{k}$~\cite{PhysRevB.83.205101, doi:10.1126/science.aaa9297, NIELSEN198120, NIELSEN1981173, doi:10.1126/science.aaa9273, doi:10.1126/science.aaq1221}, where $S^{2}$ is the closed two-dimensional (2D) sphere enclosing the Weyl point and $F$ is the Berry curvature. Moreover, there are surface Fermi arcs connecting the projections of the opposite chiral Weyl points~\cite{PhysRevB.83.205101, doi:10.1126/science.aaa9297, Xu_2015}.

Later, Lian and Zhang generalized three-dimensional (3D) Weyl semimetals to five dimensions~\cite{PhysRevB.94.041105}. According to the symmetries, a five-dimensional (5D) topological semimetal can be classified into two distinct categories: those hosting 0D Yang monopoles~\cite{10.1063/1.523506, doi:10.1126/science.aam9031, HASEBE2020115012, PhysRevLett.130.243801, PhysRevResearch.4.033203} and those hosting 2D linked Weyl surfaces~\cite{PhysRevB.94.041105, PhysRevB.95.235106, PhysRevB.95.165443, PhysRevB.100.075112, PhysRevB.101.245138, doi:10.1126/science.abi7803, PhysRevB.109.085307, PhysRevResearch.4.033203, Mathai_2017}. When the \textbf{TP} (time reversal combined with space inversion) symmetry is preserved, the conduction and valence bands intersect at Yang monopoles. The Yang monopole carries a nonzero second Chern number $C_{2}=\frac{1}{8\pi^{2}}\oint_{S^{4}}(\Omega\wedge\Omega)d^{4}\textbf{k}$~\cite{doi:10.1126/science.aam9031, PhysRevLett.130.243801, Mathai_2017, PhysRevB.100.075423}, where $S^{4}$ is the closed four-dimensional (4D) manifold enclosing the Yang monopole and $\Omega$ is the non-Abelian Berry curvature. When the \textbf{TP} symmetry is broken, the 0D Yang monopoles evolve into 2D linked Weyl surfaces, and these Weyl surfaces form a Hopf link with a topological linking number equal to the second Chern number~\cite{PhysRevB.94.041105, PhysRevB.95.235106, doi:10.1126/science.abi7803}.

Floquet topological phases, which are referred to the periodic driving induced topological states, have been widely investigated in lower dimensions~\cite{Rechtsman_2013, Lindner_2011, Wintersperger_2020, McIver_2019, PhysRevLett.107.216601, PhysRevLett.109.010601, PhysRevLett.110.200403, PhysRevLett.111.136402, PhysRevLett.112.026805, PhysRevLett.113.266801, PhysRevLett.121.237401, PhysRevX.3.031005, H_bener_2017, Rudner_2020, PhysRevResearch.6.023010, PhysRevB.79.081406, PhysRevB.82.235114, PhysRevA.77.031405, Zhan_2023, Jangjan_2020, PhysRevB.102.041119, PhysRevB.106.224306, PhysRevB.106.235405, PhysRevB.107.235132, PhysRevB.108.075435, PhysRevB.108.195125, 10.1088/1361-648X/ac530f, 10.1002/pssr.201206451, PhysRevB.110.235140, PhysRevB.109.085148, arXiv2401.18038Qin, arXiv2406.04763Zhang, arXiv2407.10191Stegmaier, arXiv2409.02774zhan}. Specifically, Floquet topological semimetals possess novel topological properties without analogs in static systems~\cite{H_bener_2017, Wang_2014, PhysRevLett.117.087402, PhysRevLett.120.237403, PhysRevLett.121.036401, PhysRevLett.121.196401, PhysRevLett.123.066403, PhysRevResearch.2.033045, PhysRevResearch.3.L032026, PhysRevB.93.144114, PhysRevB.94.075443, PhysRevB.94.235137, PhysRevB.94.121106, PhysRevB.94.235447, PhysRevB.96.041206, PhysRevB.96.041205, PhysRevB.96.041126, PhysRevB.97.155152, PhysRevB.102.201105, PhysRevB.103.094309, PhysRevB.104.205117, PhysRevB.105.L081102, PhysRevB.105.L121113, PhysRevB.108.205139, PhysRevB.107.L121407, PhysRevB.110.125204}. For example, in static systems, a Weyl semimetal must host at least one pair of Weyl points with opposite chirality~\cite{NIELSEN198120, NIELSEN1981173}, while the Floquet topological semimetal supports the existence of a single Weyl point~\cite{PhysRevLett.121.196401, PhysRevLett.123.066403}. Experimentally, Floquet topological semimetals have been  realized in various metamaterials~\cite{Rechtsman_2013, doi:10.1126/science.1239834, Jotzu_2014, Fleury_2016, Peng_2016, Maczewsky_2017, Mukherjee_2017, Asteria_2019, RevModPhys.91.015006, McIver_2019, Maczewsky_2020, Wintersperger_2020, 10.1364/PRJ.404163, 10.1063/5.0150118}. However, research on Floquet topological semimetals has been largely confined to 2D or 3D systems, and higher-dimensional Floquet topological semimetals remain unexplored.

\begin{figure*}[t]
	\includegraphics[width=0.9\textwidth]{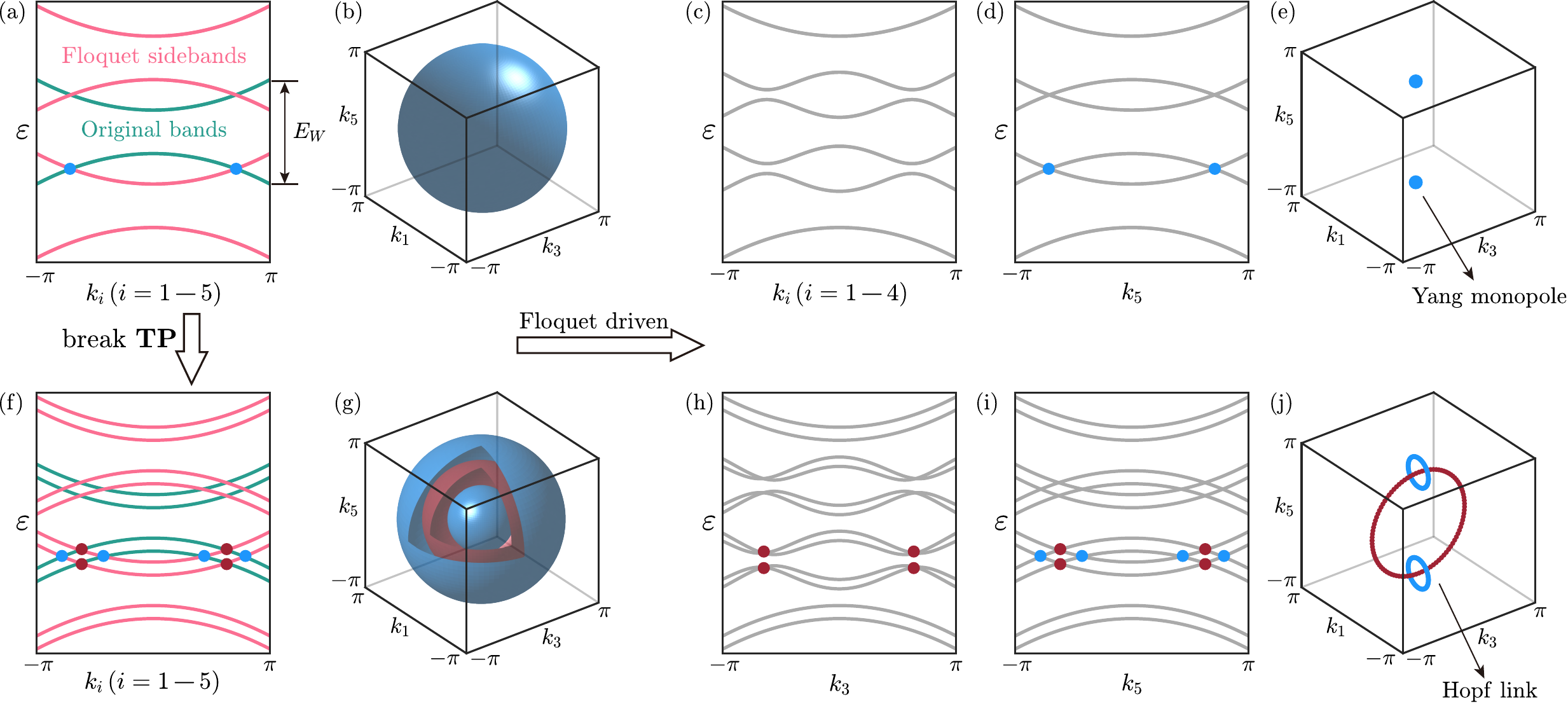} \caption{(a) The quasienergy $\varepsilon$ of the Floquet system as a function of $k_{i}$ ($i=1, 2, 3, 4, 5$) when the strength of time-periodic driving $m_{t}=0$. In the presence of the \textbf{TP} symmetry, the bulk bands are twofold degenerate. The green and pink lines indicate the original bands of $H(\textbf{k})$ and the Floquet sidebands of $H(\textbf{k})\pm\omega$, respectively, where $\omega$ is the frequency of time-periodic driving. When $\omega<E_{W}$, they intersect at the light blue points in the resonant quasienergy region, and $E_{W}$ represents the bandwidth of the static Hamiltonian $H(\textbf{k})$. (b) The light blue gapless points form a closed spherical surface in the $(k_{1}, k_{3}, k_{5})$ subspace when $(k_{2}, k_{4})$ are fixed. (c)-(e) When $m_{t}\neq0$, the Floquet sidebands couple with the original bands, resulting in the bulk gap being opened in the first Brillouin zone, except for two 0D Yang monopoles located on the $k_{5}$ axis. Similar evolutionary processes are shown in (f)-(j) when the \textbf{TP} symmetry is broken. (f) In the absence of the \textbf{TP} symmetry, the degeneracy of the bulk bands is destroyed. In the resonant quasienergy region, there are bulk gapless points marked by the light blue dots between bands 4 and 5, and by the red dots between bands 5 and 6 (or bands 3 and 4). (g) The bulk gapless points form three concentric spherical surfaces. (h)-(j) When $m_{t}\neq0$, the time-periodic driving forces the bulk gap to open, except for a Hopf link formed by linked Weyl surfaces.}%
	\label{fig1}
\end{figure*}

In this paper, we investigate the effects of a time-periodic driving on the 5D \textbf{TP} symmetry-preserving (symmetry-breaking) normal insulator. We focus on topological phase transitions in the resonant quasienergy region, where the Floquet sidebands overlap with the original bands as shown in Figs.~\ref{fig1}(a) and \ref{fig1}(f). First, we introduce the time-periodic driving to the 5D \textbf{TP} symmetry-preserving normal insulator. In the 5D momentum space, the bulk gapless points form a closed 4D manifold because of the bands overlapping in the resonant quasienergy region,  which is projected as a closed 2D sphere in the $(k_{1}, k_{3}, k_{5})$ subspace as shown in Fig.~\ref{fig1}(b). Meanwhile, the time-periodic driving induces the coupling between the Floquet sidebands and the original bands, resulting in the 4D manifold collapsing into two 0D gapless points located on the $k_{5}$ axis as shown in Figs.~\ref{fig1}(c)--(e). These two 0D gapless points are confirmed to be Yang monopoles characterized by the second Chern number. Besides, we find there exist surface Weyl arcs connecting these two 0D Yang monopoles. These features suggest that the time-periodic driving induces a 5D Floquet topological semimetal with Yang monopoles.

Second, we introduce the time-periodic driving to the 5D \textbf{TP} symmetry-breaking normal insulator. In the 5D momentum space, the 4D manifold splits into three 4D manifolds since the band degeneracy is broken. As shown in Fig.~\ref{fig1}(g), the projections of these three 4D manifolds in the $(k_{1}, k_{3}, k_{5})$ subspace appear as three concentric spheres. Meanwhile, the time-periodic driving forces the three 4D manifolds to shrink into the 2D linked Weyl surfaces. In Fig.~\ref{fig1}(j), the red and light blue loops are the projections of the Weyl surfaces shown in the $(k_{1}, k_{3}, k_{5})$ subspace. These linked Weyl surfaces form a Hopf link with the topological linking number equal to the second Chern number, which is equal to the number of surface Weyl arcs. Additionally, both the two 5D Floquet topological semimetals transition into 5D Floquet Chern insulators by further increasing the strength of the time-periodic driving.

The paper is organized as follows. In Sec.~\ref{SecII}, a 5D time-dependent model and the method for calculating the second Chern number are presented. In Sec.~\ref{SecIII}, numerical calculations show that the effect of the time-periodic driving on the \textbf{TP} symmetry-preserving normal insulator, and a 5D Floquet topological semimetal with Yang monopoles is proposed. In Sec.~\ref{SecIV}, a 5D Floquet topological semimetal with linked Weyl surfaces induced by the time-periodic driving is proposed. In Sec.~\ref{Conclusion}, we summarize our results.
\vspace{-0.1cm}
\section{Model and Method}
\label{SecII}
The time-dependent 5D topological semimetal model is given by
\begin{align}
H(\textbf{k},t)=H(\textbf{k})+M(t).
\label{eq1}
\end{align}
The first term in Eq.~(\ref{eq1}) represents a static Hamiltonian describing the 5D topological semimetal~\cite{PhysRevB.95.235106},
\begin{align}
H(\textbf{k})=\sum_{j=1}^{5}\xi_{j}(\textbf{k})\gamma_{j}+b\frac{i[\gamma_{3}, \gamma_{4}]}{2},
\label{eq2}
\end{align}
where $\xi_{j}(\textbf{k})=\sin(k_{j})$ for $j=1, 2, 3, 4$ and $\xi_{5}(\textbf{k})=m+\sum_{j=1}^{4}[1-\cos(k_{j})]+\eta [1-\cos(k_{5})]$. The gamma matrices are written as $\gamma_{1}=\sigma_{3}\otimes\sigma_{1}$, $\gamma_{2}=\sigma_{3}\otimes\sigma_{2}$, $\gamma_{3}=\sigma_{3}\otimes\sigma_{3}$, $\gamma_{4}=\sigma_{1}\otimes\sigma_{0}$, and $\gamma_{5}=\sigma_{2}\otimes\sigma_{0}$. $\sigma_{0}$ and $\sigma_{j}$ $(j=1, 2, 3)$ are identity matrix and Pauli matrices, respectively. $m$ and $\eta$ are the tuning parameters. In the subsequent calculations, $\eta=0.5$. The second term in Eq.~(\ref{eq1}) represents the time-periodic on-site potential $M(t)=m_{t}\cos(\omega t)\gamma_{5}$, where $m_{t}$ and $\omega$ are the strength and frequency of the time-periodic on-site potential, respectively.

The first term in Eq.~(\ref{eq2}) describes a 5D topological semimetal with the \textbf{TP} symmetry $\mathcal{TP}=\sigma_{1}\otimes\sigma_{2}\mathcal{K}$, where time-reversal symmetry $\mathcal{T}=i\gamma_{2}\mathcal{K}$ and space-inversion symmetry $\mathcal{P}=\gamma_{5}$ satisfy $\mathcal{T}H(\textbf{k})\mathcal{T}^{-1}=H(-\textbf{k})$ and $\mathcal{P}H(\textbf{k})\mathcal{P}^{-1}=H(-\textbf{k})$, respectively. When the second term $b \neq 0$ in Eq.~(\ref{eq2}), the \textbf{TP} symmetry is broken. When $m>b$, the static system is a 5D normal insulator~\cite{PhysRevB.95.235106}.

In 3D Weyl semimetals, there are pairs of Weyl points that carry topological monopole charges given by the first Chern number in the momentum space~\cite{NIELSEN198120}, and the 2D planes in the 3D Brillouin zone that lie between the pair of Weyl points have nonzero first Chern numbers~\cite{10.1146/annurev-conmatphys-031113-133841}. Moreover, there are surface Fermi arcs connecting the projections of the oppositely charged Weyl points~\cite{PhysRevB.83.205101}. Similar to the 3D Weyl semimetal, the 5D topological semimetal hosts topologically protected Weyl arcs on its 4D boundaries, and the arcs connect the projections of the two Yang monopoles or Weyl surfaces~\cite{PhysRevB.94.041105, PhysRevB.95.235106}. The Yang monopole or the Weyl surface is characterized by a nonzero second Chern number $C_{2}$, and the second Chern number in a 4D momentum space ($k_{1}, k_{2}, k_{3}, k_{4}$) is given by~\cite{PhysRevB.78.195424, Mochol_Grzelak_2018, doi:10.1126/science.aam9031, PhysRevB.108.085306, PRXQuantum.2.010310, PhysRevB.108.085114, PhysRevB.109.125303, 10.1088/0256-307X/41/4/047102, PhysRevB.110.195144}
\begin{align}
C_{2}=\frac{1}{4\pi^{2}}\int_{\mathcal{V}} d^{4}\textbf{k}\text{Tr}[\Omega_{12}\Omega_{34}+\Omega_{41}\Omega_{32}+\Omega_{31}\Omega_{24}],
\end{align}
where $\mathcal{V}$ represents the first Brillouin zone of the 4D momentum space. The non-Abelian Berry curvature is
\begin{align}
\Omega_{pq}^{\alpha\beta}=\partial_{p}A_{q}^{\alpha\beta}-\partial_{q}A_{p}^{\alpha\beta}+i[A_{p},A_{q}]^{\alpha\beta},
\end{align}
where $p, q=1, 2, 3, 4$, and the Berry connection of the occupied bands is
\begin{align}
A_{p}^{\alpha\beta}=&-i\left\langle u^{\alpha}(\textbf{k})\right\vert\frac{\partial}{\partial k_{p}}\left\vert u^{\beta}(\textbf{k})\right\rangle.
\end{align}
$\left\vert u^{\alpha}(\textbf{k})\right\rangle$ denotes the occupied eigenstates below the Fermi energy $\varepsilon_{F}$ with $\alpha=1, \dots, N_{\rm{occ}}$.

\section{5D Floquet topological semimetal with Yang monopoles}
\label{SecIII}
\begin{figure}[t]
	\includegraphics[width=0.48\textwidth]{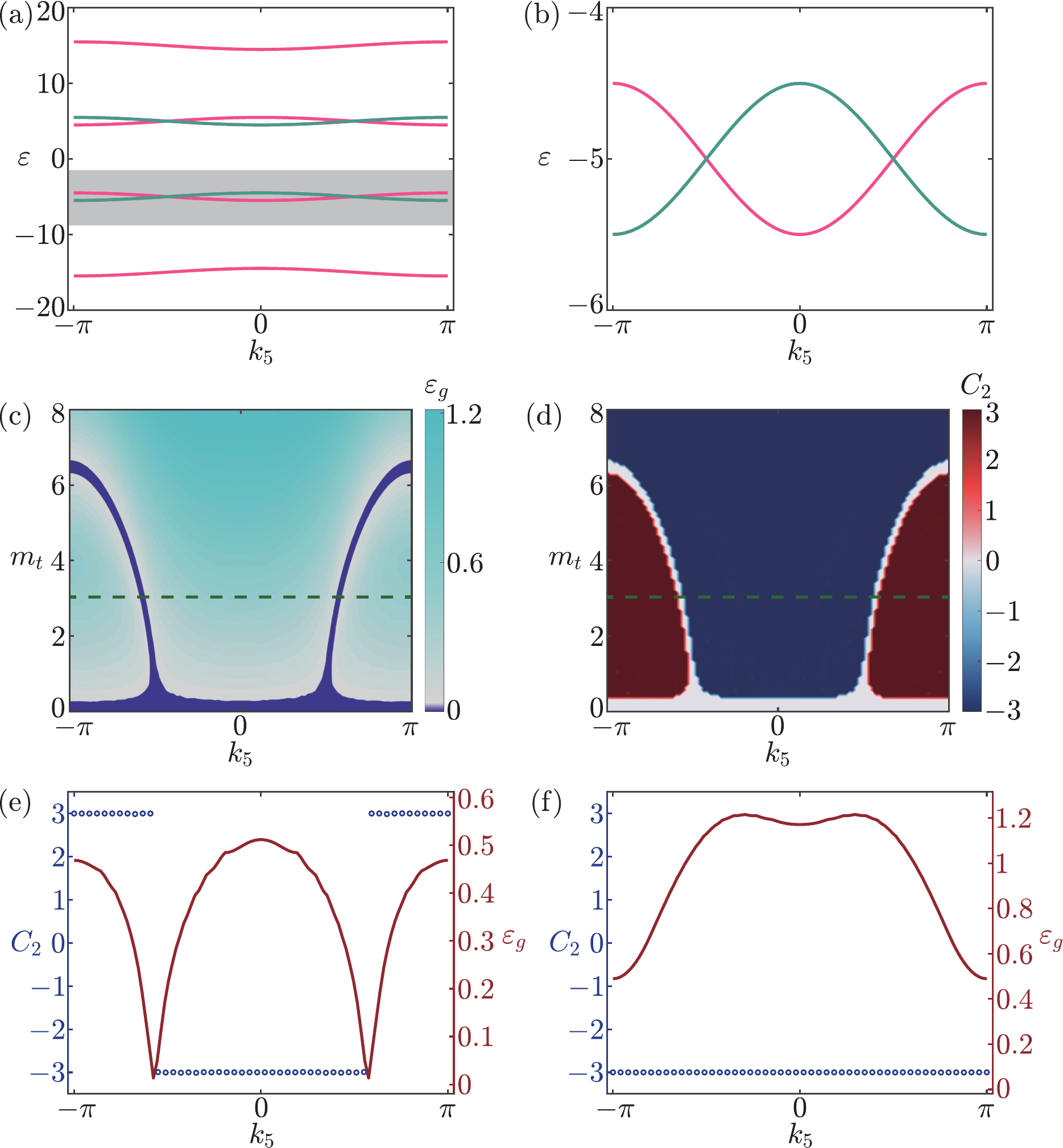} \caption{Topological phases in the Floquet system with the \textbf{TP} symmetry. (a) The bulk quasienergy $\varepsilon$ as a function of $k_{5}$ when $m_{t}=0$ and $(k_{1}, k_{2}, k_{3}, k_{4})=(\pi, \pi, 0, 0)$. The green (pink) line represents the band of the diagonal block $H_{0}$ ($H_{0}\pm\omega$) in the Floquet Hamiltonian $H_{F}$. The gray region is the resonant quasienergy region we are concerned with. (b) The quasienergy spectra of the gray region in (a). (c)~The bulk gap $\varepsilon_{g}$ in the plane $(m_{t}, k_{5})$ when $N_{\rm{occ}}=4$. (d) The second Chern number $C_{2}$ in the plane $(m_{t}, k_{5})$ when $N_{\rm{occ}}=4$. (e)--(f) The $k_{5}$-dependent second Chern number $C_{2}$ (blue circles) and the bulk gap $\varepsilon_{g}$ (red line). The amplitude of the time-periodic driving for (e) and (f) is $m_{t}=3$ and $m_{t}=8$, respectively. Here, we choose $\eta=0.5$, $m=0.5$, $b=0$, and $\omega=10$.}%
	\label{fig2}
\end{figure}

In this section, we explore the topological phase transition of the 5D Floquet system by applying a time-periodic driving to the \textbf{TP} symmetry-preserving normal insulator. Using Floquet theory~\cite{PhysRev.138.B979}, the time-dependent Hamiltonian $H(\textbf{k},t)$ in Eq.~(\ref{eq1}) can be expressed as the Floquet Hamiltonian $H_{F}$ through Fourier transformation. The Floquet Hamiltonian $H_{F}$ is given by
\begin{align}
H_{F}=\begin{pmatrix}
\ddots & \vdots & \vdots & \vdots & \ddots \\
\cdots & H_{-1,-1}-\omega & H_{-1,0} & H_{-1,1} & \cdots \\
\cdots & H_{0,-1} & H_{0,0} & H_{0,1} & \cdots \\
\cdots & H_{1,-1} & H_{1,0} & H_{1,1}+\omega  & \cdots \\
\ddots & \vdots & \vdots & \vdots & \ddots
\end{pmatrix},
\end{align}
where
\begin{align}
H_{n,n'}&=\frac{1}{T}\int_{0}^{T} H(\textbf{k},t)e^{i (n'-n)\omega t}d t
\end{align}
with $n,n'=0,\pm1,\pm2,\cdots$ is the harmonic component of the time-dependent Hamiltonian $H(\textbf{k},t)$, and $T=\frac{2\pi}{\omega}$. In this paper, $H_{n,n}=H(\textbf{k})$, $H_{n,n\pm1}=\frac{m_{t}}{2}\gamma_{5}$, and $H_{n,n\pm l}=0$ ($l\geq2$). Then, we can obtain a $(2n_{t}+1)N_{\rm{band}}\times (2n_{t}+1)N_{\rm{band}}$ effective Floquet Hamiltonian by truncating the Floquet space. $N_{\rm{band}}=4$ is the number of energy bands of the static system. $n_{t}$ is a positive integer, representing the truncation number~\cite{PhysRevB.109.224315}, and $n,n'\in [-n_{t}, n_{t}]$. In subsequent calculations, we select the truncation number $n_{t}=1$.

\begin{figure*}[t]
	\includegraphics[width=0.98\textwidth]{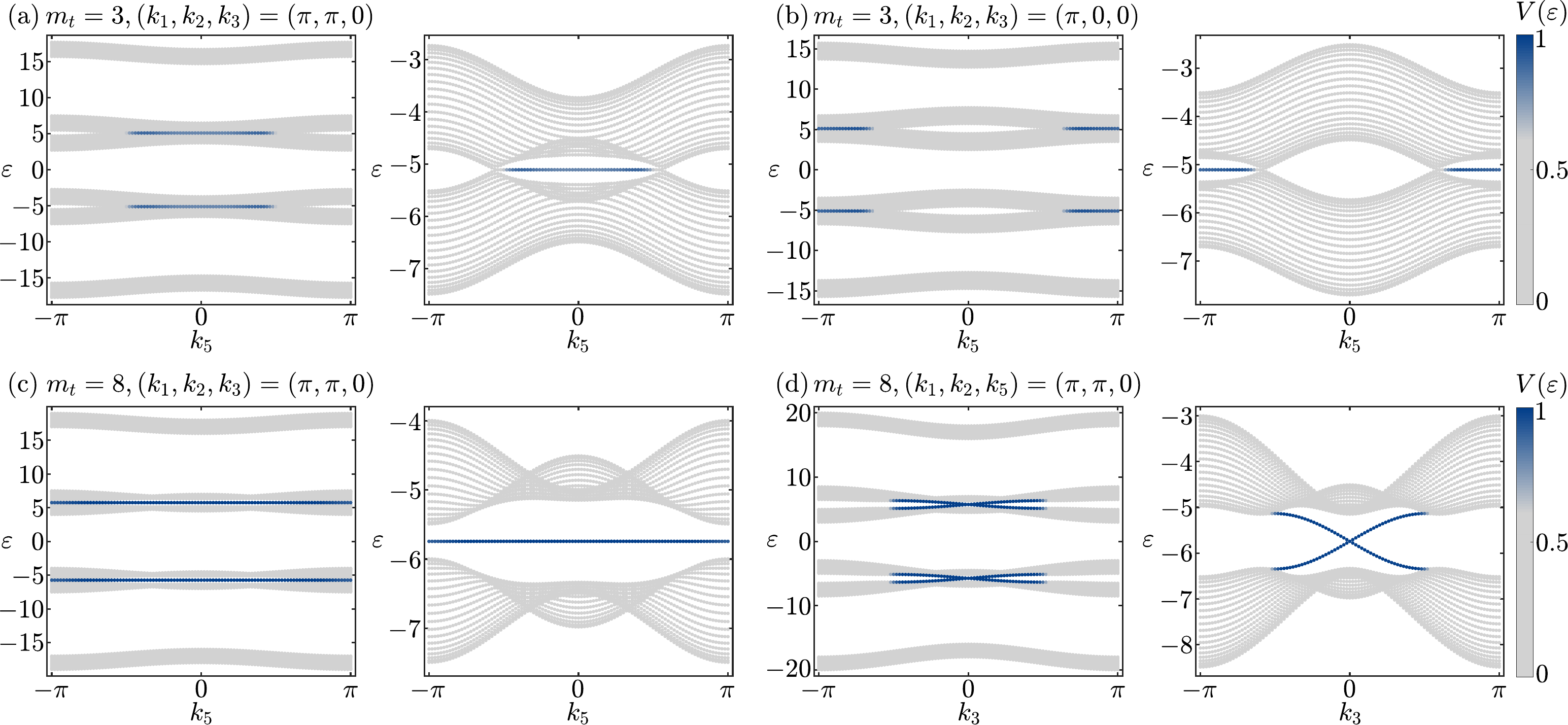} \caption{Quasienergy spectra of the Floquet system when the open boundary condition is along the $k_{4}$ direction. (a) The quasienergy spectra as a function of $k_{5}$ in the left panel when $m_{t}=3$ and $(k_{1}, k_{2}, k_{3})=(\pi, \pi, 0)$. (b) The quasienergy spectra as a function of $k_{5}$ in the left panel when $m_{t}=3$ and $(k_{1}, k_{2}, k_{3})=(\pi, 0, 0)$. (c) The quasienergy spectra as a function of $k_{5}$ in the left panel when $m_{t}=8$ and $(k_{1}, k_{2}, k_{3})=(\pi, \pi, 0)$. (d) The quasienergy spectra as a function of $k_{3}$ in the left panel when $m_{t}=8$ and $(k_{1}, k_{2}, k_{5})=(\pi, \pi, 0)$. In all plots, the right panel shows the spectra within the resonant quasienergy region of the left panel. The color bar corresponds to the quantity $V(\varepsilon)=\sum_{j\in N_{\rm{edge}}}|\phi_{j}(\varepsilon)|^{2}/\sum_{j\in N_{\rm{total}}}|\phi_{j}(\varepsilon)|^{2}$ for the localization degree at the boundary, where $\phi_{j}(\varepsilon)$ is the element of the eigenstate with the quasienergy $\varepsilon$ at site $j$. Here, we choose $\eta=0.5$, $m=0.5$, $b=0$, and $\omega=10$.}%
	\label{fig3}
\end{figure*}

The frequency of the time-periodic driving is in the interval $[E_{W}/2, E_{W}]$, where $E_{W}=2(m+b)+18$ represents the band width of the static system. When $E_{W}/2<\omega<E_{W}$, the original bands of $H_{0,0}$ and the Floquet sidebands of $H_{n,n}\pm\omega$ ($n=\pm1$) overlap each other as shown in Fig.~\ref{fig2}(a). By modulating the strength of the time-periodic driving $m_{t}$, we study the topological phase transition in the resonant quasienergy region [the gray region in Fig.~\ref{fig2}(a)].

In Fig.~\ref{fig2}(c), we show the bulk gap $\varepsilon_{g}$ as a function of $m_{t}$ and $k_{5}$, with the green dashed line given by $m_{t}=3$. The dark blue region in the color bar represents the bulk gap tending to zero. When the strength of the time-periodic driving is weak, the bulk band is gapless for any $k_{5}$. When $m_{t}$ is increased to a suitable strength, there are only two gapless points in the first Brillouin zone. Using $k_{5}$ as a parameter, we show the evolution of the second Chern number $C_{2}$ on the $(k_{5}, m_{t})$ plane as shown in Fig.~\ref{fig2}(d). Compared with Fig.~\ref{fig2}(c), it can be found that the second Chern number exhibits a nonzero integer in the gapped regions. In Fig.~\ref{fig2}(e), we show the changes in the bulk gap $\varepsilon_{g}$ and the second Chern number $C_{2}$ with respect to $k_{5}$ when $m_{t}=3$. It can be observed that there are only two gapless points located on the $k_{5}$ axis. With the gapless points as boundaries, the second Chern number near $k_{5}=0$ manifests as a quantized integer $C_{2}=-3$, while the second Chern number near $k_{5}=\pm\pi$ is $C_{2}=3$. In Fig.~\ref{fig3}, we show that the quasienergy spectra of the Floquet system when the open boundary condition is along the $k_{4}$ direction. When $m_{t}=3$, there are surface Weyl arcs connecting the projections of the bulk gapless points as shown in Figs.~\ref{fig3}(a) and \ref{fig3}(b). These features suggest that the time-periodic driving induces a 5D Floquet topological semimetal, with the bulk gapless points being Yang monopoles. In the region near $k_{5}=0$, the surface Weyl arcs are distributed at $(k_{1}, k_{2}, k_{3})=(\pi, \pi, 0)$, $(k_{1}, k_{2}, k_{3})=(\pi, 0, \pi)$, and $(k_{1}, k_{2}, k_{3})=(0, \pi, \pi)$, while in the region near $k_{5}=\pm\pi$, the surface Weyl arcs are distributed at $(k_{1}, k_{2}, k_{3})=(\pi, 0, 0)$, $(k_{1}, k_{2}, k_{3})=(0, \pi, 0)$, and $(k_{1}, k_{2}, k_{3})=(0, 0, \pi)$. The number and positions of the surface Weyl arcs are related to the value and sign of the second Chern number.

By further increasing the strength of the time-periodic driving, the bulk gap is opened. In Fig.~\ref{fig2}(f), we show the bulk gap $\varepsilon_{g}$ and the second Chern number $C_{2}$ as functions of $k_{5}$ when $m_{t}=8$. For any $k_{5}$, the second Chern number is always $C_{2}=-3$. Clearly, this bulk gap opened by the time-periodic driving is topologically nontrivial. Figures \ref{fig3}(c) and \ref{fig3}(d) respectively show the quasienergy spectra as a function of $k_{5}$ and $k_{3}$ with open boundary conditions along the $k_{4}$ direction. In Fig.~\ref{fig3}(c), it can be observed that when $(k_{1}, k_{2}, k_{3})=(\pi, \pi, 0)$, there is a flat band along the $k_{5}$ axis within the bulk gap. Notably, there are also flat boundary states along the $k_{5}$ axis when $(k_{1}, k_{2}, k_{3})=(\pi, 0, \pi)$ or $(k_{1}, k_{2}, k_{3})=(0, \pi, \pi)$. In Fig.~\ref{fig3}(d), there are gapless boundary states in the bulk gap when $(k_{1}, k_{2}, k_{5})=(\pi, \pi, 0)$. Combining the second Chern number and the quasienergy spectra, we demonstrate that this system is a 5D Floquet Chern insulator characterized by $C_{2}=-3$.

\section{5D Floquet topological semimetal with Linked Weyl surfaces}
\label{SecIV}
\begin{figure}[t]
	\includegraphics[width=0.48\textwidth]{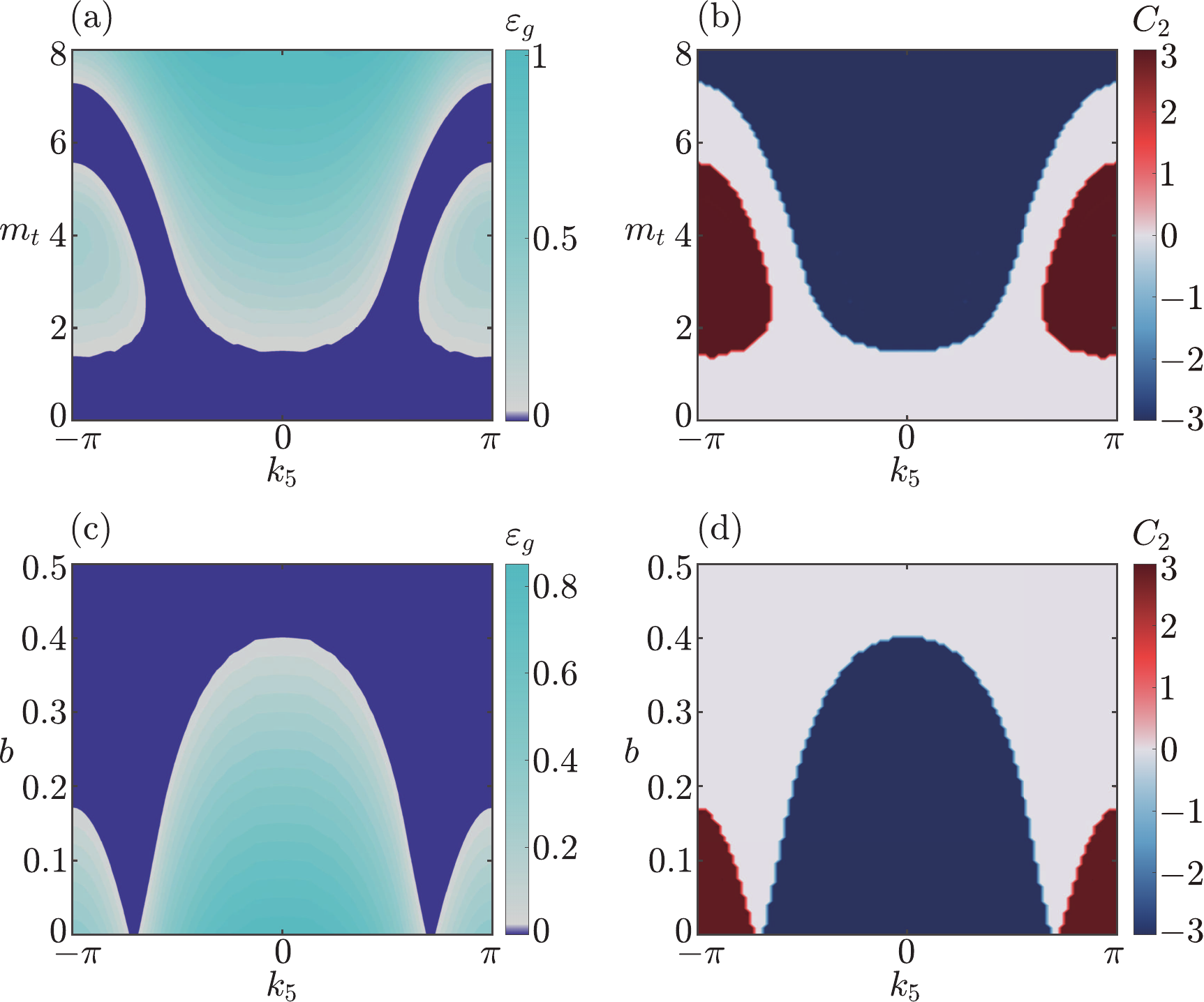} \caption{Phase diagram in the \textbf{TP} symmetry-breaking Floquet system. (a)~The bulk gap $\varepsilon_{g}$ in the plane $(m_{t}, k_{5})$. (b) The second Chern number $C_{2}$ in the plane $(m_{t}, k_{5})$. In (a) and (b), $b=0.1$. (c)~The bulk gap $\varepsilon_{g}$ in the plane $(b, k_{5})$. (d) The second Chern number $C_{2}$ in the plane $(b, k_{5})$. In (c) and (d), $m_{t}=5$. Here, we choose the number of the occupied states $N_{\rm{occ}}=4$, $\eta=0.5$ and $\omega=10$.}%
	\label{fig4}
\end{figure}

In this section, we investigate the effects of the time-periodic driving on the 5D \textbf{TP} symmetry-breaking normal insulator. In Fig.~\ref{fig4}, we present the evolution of the bulk gap $\varepsilon_{g}$ and the second Chern number $C_{2}$ in the plane $(m_{t}, k_{5})$ and the plane $(b, k_{5})$, aiming to study the effect of the strength of the time-periodic driving on the \textbf{TP} symmetry-breaking system, as well as the effect of the \textbf{TP} symmetry-breaking term on the 5D Floquet topological semimetal.

\begin{figure}[t]
	\includegraphics[width=0.48\textwidth]{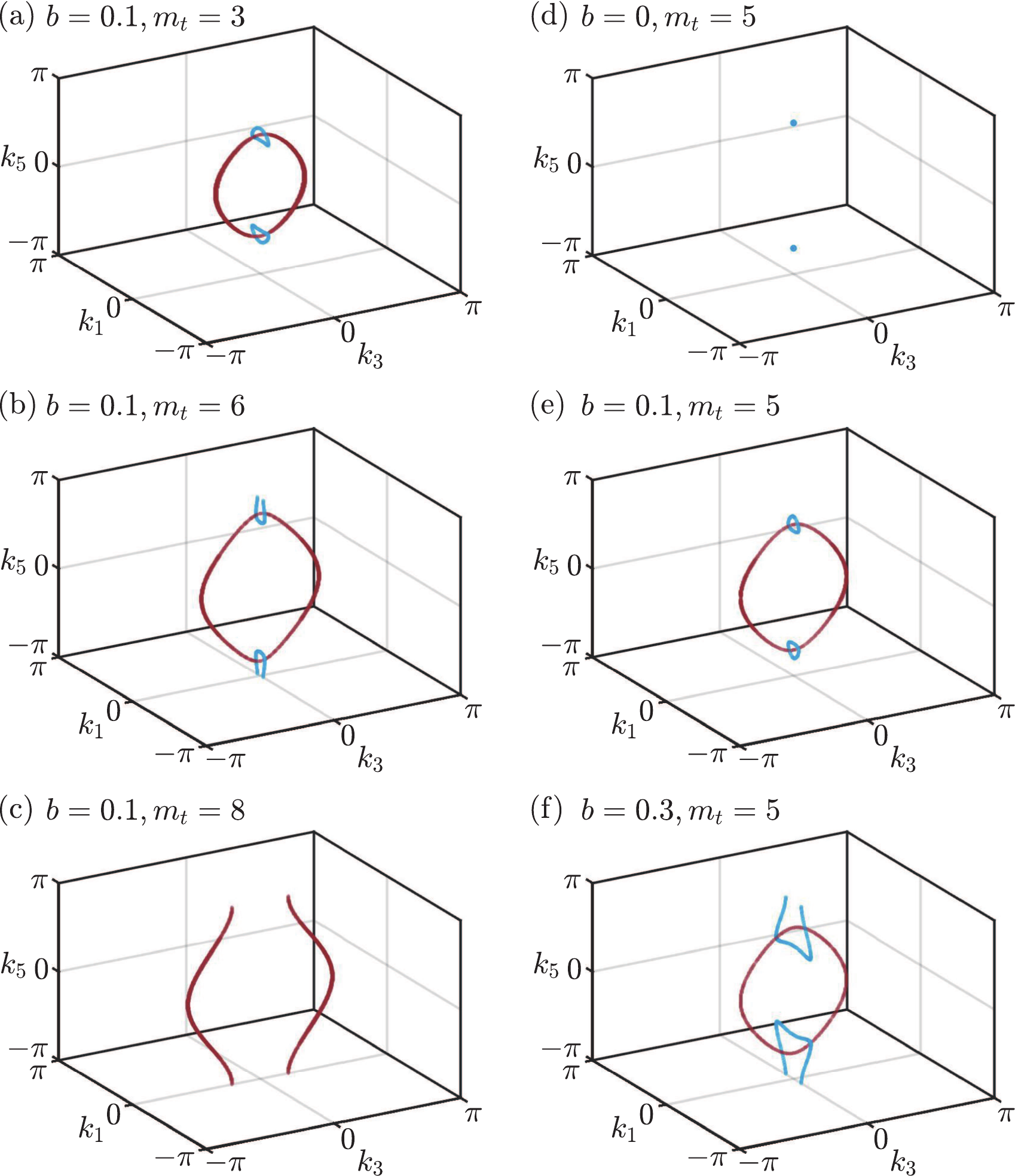} \caption{The distribution of bulk gapless points in the $(k_{1}, k_{3}, k_{5})$ subspace for (a)~$b=0.1, m_{t}=3$; (b)~$b=0.1, m_{t}=6$; (c)~$b=0.1, m_{t}=8$; (d)~$b=0, m_{t}=5$; (e)~$b=0.1, m_{t}=5$; and (f)~$b=0.3, m_{t}=5$. In (a)--(f), $k_{2}=k_{4}=\pi$. The red (light blue) loops represent the Weyl surfaces between bands 5 and 6 (between bands 4 and 5). Here, we choose $\eta=0.5$ and $\omega=10$.}%
	\label{fig5}
\end{figure}

When $E_{W}/2<\omega<E_{W}$, the original bands of $H_{0,0}$ and the Floquet sidebands of $H_{n,n}\pm\omega$ ($n=\pm1$) overlap each other, resulting in the bulk gap $\varepsilon_{g}=0$ when $m_{t}=0$ and $N_{\rm{occ}}=4$ as shown in Fig.~\ref{fig4}(a). When $m_{t}=3$ and $(k_{2}, k_{4})=(\pi, \pi)$, there are two Weyl surfaces (light blue loops) distributed along the $k_{5}$ axis, as shown in Fig.~\ref{fig5}(a). Meanwhile, there is another Weyl surface (red loops) in the $(k_{1}, k_{3}, k_{5})$ subspace for $N_{\rm{occ}}=5$, which links with the light blue Weyl surfaces. Obviously, when $(k_{2}, k_{4})=(\pi, \pi)$, a single Hopf link appears in the $(k_{1}, k_{3}, k_{5})$ subspace near $k_{5}=0$, as shown in Fig.~\ref{fig5}(a). By varying $k_{2}$ and $k_{4}$, we identify six Hopf links in the full 5D momentum space, with three located near $k_{5}=0$, and the remaining three near $k_{5}=\pm\pi$. Consequently, the linking numbers of the Weyl surfaces for $N_{\rm{occ}}=4$ in both the $k_{5}=0$ and $k_{5}=\pm\pi$ regions are equal to three. A comparison with Fig.~\ref{fig4}(b) reveals that the linking number of the Weyl surfaces corresponds to the value of the second Chern number, with opposite signs reflecting distinct Hopf link distributions. We thus characterize this system as a 5D Floquet topological semimetal with linked Weyl surfaces. Moreover, with increasing $m_{t}$, the two gapless regions move along the $+k_{5}$ direction and $-k_{5}$ directions, respectively, until they eventually coalesce into a single gapless region near $k_{5}=\pm\pi$ as shown in Fig.~\ref{fig4}(a). This indicates that the two light blue Weyl surfaces merge into a single Weyl surface, as shown in Fig.~\ref{fig5}(b). When $m_{t}=8$, the bulk gap is opened for $N_{\rm{occ}}=4$, while for $N_{\rm{occ}}=5$, the closed Weyl surface transforms into a cylindrical Weyl surface, as shown in Fig.~\ref{fig5}(c). Then, the system becomes a 5D Floquet Chern insulator with $C_{2}=-3$.

On the other hand, we find that the \textbf{TP} symmetry-breaking term can induce phase transitions in the Floquet system. Figures~\ref{fig4}(c) and \ref{fig4}(d) show the bulk gap and the second Chern number as functions of $k_{5}$ and $b$. When $b=0$, it can be found that there are two bulk gapless points in the 5D first Brillouin zone as shown in Fig.~\ref{fig5}(d). Along the $k_{5}$ axis, the second Chern number between these two gapless points exhibits two different nonzero integers in the regions near $k_{5}=0$ and $k_{5}=\pm\pi$, indicating that the system is a Floquet topological semimetal with two Yang monopoles as shown in Sec.~\ref{SecIII}. When $b>0$, the 0D Yang monopoles extend into 2D closed Weyl surfaces (light blue loops), as shown in Fig.~\ref{fig5}(e), thereby transforming the Floquet topological semimetal with Yang monopoles into a Floquet topological semimetal with linked Weyl surfaces. As $b$ increases further,  the two Weyl surfaces move closer together along the $k_{5}$ axis until they merge into one as shown in Fig.~\ref{fig5}(f). When $b$ exceeds a critical value, the bulk gap completely closes for all $k_{5}$, and the system transitions into a trivial metal phase.

\section{Conclusions}
\label{Conclusion}
In this paper, we investigate the effects of a on-site time-periodic driving on the 5D \textbf{TP} symmetry-preserving (symmetry-breaking) normal insulator. First, we investigate the effects of the on-site time-periodic driving on the 5D \textbf{TP} symmetry-preserving normal insulator. Our calculations show that the time-periodic driving can induce a topological phase transition from the normal insulator to a 5D Floquet topological semimetal with Yang monopoles. We find that the second Chern number between the two Yang monopoles is a nonzero integer, with its magnitude and sign reflecting the number and position of the surface Weyl arcs, respectively. As the strength of the time-periodic driving increases, the two Yang monopoles gradually approach and eventually annihilate. Then the bulk gap reopens, and the system transitions from the 5D Floquet topological semimetal to a 5D Floquet Chern insulator characterized by $C_{2}=-3$.

Second, we investigate the effects of the time-periodic driving on the 5D \textbf{TP} symmetry-breaking normal insulator. We find that the time-periodic driving can induce a transition from the normal insulator to a Floquet topological semimetal with linked Weyl surfaces, and even to a Floquet Chern insulator characterized by $C_{2}=-3$. Moreover, we observe that the \textbf{TP} symmetry-breaking term can modulate the topological phases of the 5D Floquet topological semimetal. When the \textbf{TP} symmetry-breaking term is weak, it drives the transition from the 5D Floquet topological semimetal hosting Yang monopoles to one hosting linked Weyl surfaces. Furthermore, the stronger \textbf{TP} symmetry-breaking term can disrupt the Floquet topological phase, resulting in a transition to a trivial metal phase.

Experimentally, 5D Weyl semimetals have been realized in artificial systems such as photonic metamaterials~\cite{doi:10.1126/science.abi7803}, microwaves~\cite{PhysRevLett.130.243801}, and electric circuits~\cite{PhysRevResearch.4.033203, PhysRevB.109.085307}. Moreover, recent experiments have shown that artificial metamaterials, such as photonic crystals~\cite{Rechtsman_2013, Mukherjee_2017, Maczewsky_2017, RevModPhys.91.015006, Maczewsky_2020} and electric circuits~\cite{10.1063/5.0150118}, are ideal platforms for realizing Floquet topological matters. Therefore, it is expected that the 5D Floquet topological semimetals can be realized based on the artificial metamaterials.

\section*{Acknowledgments}
B.Z. was supported by the NSFC (Grant No. 12074107), the program of outstanding young and middle-aged scientific and technological innovation team of colleges and universities in Hubei Province (Grant No. T2020001) and the innovation group project of the Natural Science Foundation of Hubei Province of China (Grant No. 2022CFA012). R.C. acknowledges the support of NSFC (under Grant No. 12304195) and the Chutian Scholars Program in Hubei Province. Z.-R.L. was supported by the National Funded Postdoctoral Researcher Program (under Grant No. GZC20230751) and the Postdoctoral Innovation Research Program in Hubei Province (under Grant No. 351342).

%\newpage
\bibliographystyle{apsrev4-1-etal-title_6authors}

\end{document}